\documentclass[useAMS,usenatbib,usegraphicx,letterpaper]{mn2e}
\usepackage{times,amssymb,amsmath,hyperref,aas_macros}

\usepackage[totalwidth=480pt,totalheight=680pt,layoutvoffset=0.5cm]{geometry}

\begin{document}

\title[Exocomet Orbit Fitting]{Exocomet Orbit Fitting: Accelerating Coma
  Absorption During Transits of $\beta$ Pictoris}

\author[Grant M. Kennedy]{Grant M. Kennedy\thanks{Email:
    \href{mailto:g.kennedy@warwick.ac.uk}{g.kennedy@warwick.ac.uk}}$^{1,2}$ \\
  $^1$ Department of Physics, University of Warwick, Gibbet Hill Road,
  Coventry, CV4 7AL, UK \\
  $^2$ Centre for Exoplanets and Habitability, University of Warwick, Gibbet Hill Road,
  Coventry, CV4 7AL, UK 
}

\maketitle

\begin{abstract}
  Comets are a remarkable feature in our night sky, visible on their
  passage through the inner Solar system as the Sun's energy sublimates
  ices and liberates surface material, generating beautiful comae, dust,
  and ion tails. Comets are also thought to orbit other stars, and are
  the most promising interpretation of sporadic absorption features
  (i.e. transits) seen in spectra of stars such as $\beta$~Pictoris and
  49~Ceti. These ``exocomets'' are thought to form and evolve in the
  same way as in the Solar system, and as in the Solar system we may
  gain insight into their origins by deriving their orbits. In the case
  of $\beta$ Pictoris, orbits have been estimated indirectly, using the
  radial velocity of the absorption features coupled with a physical
  evaporation model to estimate the stellocentric distance at transit
  $d_{\rm tr}$. Here, we note that the inferred $d_{\rm tr}$ imply that
  some absorption signatures should accelerate over several hours, and
  show that this acceleration is indeed seen in HARPS spectra. This new
  constraint means that orbital characteristics can be obtained
  directly, and the pericentre distance and longitude constrained when
  parabolic orbits are assumed. The results from fitting orbits to 12
  accelerating features, and a handful of non-accelerating ones, are in
  broad agreement with previous estimates based on an evaporation model,
  thereby providing some validation of the exocomet hypothesis. A
  prediction of the evaporation model, that coma absorption is deeper
  for more distant transits, is also seen here.
\end{abstract}

\begin{keywords}
  comets: general --- planets and satellites: detection --- planetary
  systems --- planet-disc interactions --- circumstellar matter ---
  stars: individual: $\beta$ Pictoris
\end{keywords}

\section{Introduction}\label{s:intro}

Comets are a well-known and important component of our Solar system, and
while a scientific focus for astronomers, some are visible to the naked
eye and capture the imagination of millions. These icy bodies become
visibly more active as they near the Sun, providing a window into the
primordial conditions in the outer Solar system
\citep[e.g.][]{2017MNRAS.469S.755B}. Comets formed beyond the Asteroid
belt, and are only brought within a few astronomical units through
interaction with the giant planets; in the case of shorter period comets
(e.g. Jupiter Family Comets) these interactions are recent and ongoing
\citep[e.g.][]{1997Sci...276.1670D}, while in the case of long period
comets, these interactions occurred billions of years ago and resulted
in the formation of the Oort cloud, the source of long period comets
\citep[e.g.][]{1987AJ.....94.1330D}.

While our detailed knowledge of planetary systems around other stars is
relatively limited in comparison, it is clear that processes analogous
to those postulated for the Solar system's formation and evolution do
occur. Most obviously, planets exist in spectacular abundance
\citep[e.g.][]{2013PNAS..11019273P}, and the cometary source regions
known as ``debris discs'' are seen around at least 30\% of other stars
\citep[e.g.][]{2013A&A...555A..11E,2018MNRAS.475.3046S}. Of particular
relevance here is the fact that the first clear evidence of another
planetary system -- the image of the edge-on disc orbiting $\beta$
Pictoris \citep{1984Sci...226.1421S} -- prompted a series of papers that
may mark the discovery of extrasolar comets
\citep[e.g.][]{1985ApJ...291L...1K,1987A&A...173..289L,1987A&A...185..267F,1988A&A...190..275L}. These
``exocomets'' manifested as transient red-shifted ionic absorption lines
in high-resolution ultraviolet and optical spectra; the signature of a
large coma transiting the face of the star. The exocomet interpretation
was (and perhaps still is) controversial, so these events have also
become known by the more prosaic term ``falling evaporating bodies'', or
FEBs.

These discoveries were backed up by the development of a physical model,
whereby material is ejected from a cometary nucleus into a coma
dominated by neutral hydrogen. Observed elements such as calcium and
magnesium are ionised when visible to stellar radiation, and are
subsequently blown into a tenuous ion tail by radiation pressure, thus
forming a parabolic front of metallic ions around the nucleus
\citep{1989A&A...223..304B,1990A&A...236..202B}. The effect of radiation
pressure was found to be critical to explain the observations; the
closer the coma to the star, the stronger the radiation pressure and the
smaller the front of ions, and therefore the shallower the
absorption. While subject to systematic uncertainties from model
assumptions, this property means that the distance of the coma to the
star for a given transit could be estimated from the depth of the
absorption line. The key modelling conclusion was that the velocity of
the observed absorption corresponds to the radial velocity of the coma
\citep{1990A&A...236..202B}, meaning that by measuring these velocities,
and estimating the distance to the star at transit, one may derive
constraints on the comet orbit.

A common inference from application of the model to the data for $\beta$
Pictoris, and indeed which could be inferred from the data themselves,
is that the tendency of the FEBs to be redshifted implies that the
cometary orbits are not randomly distributed. The relative lack of blue
shifted events \citep{1992A&A...264..637L,1998MNRAS.294L..31C} means
that most comets are approaching pericentre, with values for the
longitude of pericentre $\varpi$ (measured from the line of sight to the
star), being between -50 and 130$^\circ$
\citep{1990A&A...236..202B,2014Natur.514..462K}.

This conclusion led to proposals for three main theories for the origins
of $\beta$ Pictoris' transiting comets. The first is that the events
represent a single family of comets on common orbits, the aftermath of
the breakup of a single larger object \citep{1994A&A...282..804B}. The
main issue with this theory is that we are unlikely to witness the
aftermath of such an event, suggesting that mechanisms that work on
longer timescales would provide more satisfactory explanations.

One such theory appeals to the Solar system, noting that the
eccentricities of Asteroids whose pericentres precess at the same rate
as Saturn (i.e. those in the $\nu_6$ secular resonance) increase in a
way that is correlated with Saturn's pericentre. Thus, these bodies make
their closest approaches to the Sun with a preferred pericentre
direction, and the same explanation was offered for $\beta$ Pictoris'
FEBs \citep{1994Natur.372..441L}. The criticism levelled at this theory
is that it is very specific, and has not been shown to generically
reproduce the FEB phenomenon \citep[see][]{1996Icar..120..358B}.

The final theory, and the one that has received the most attention, is
that the 4:1 and/or 3:1 resonance with an outer planet lies within a
source belt
\citep{1996Icar..120..358B,2000AJ....119..397Q,2000Icar..143..170B,2001A&A...376..621T,2017A&A...605A..23P}. The
attraction of this theory is that objects in these resonances can be
excited to star- grazing orbits with a very specific range of pericentre
longitudes. A potential weakness of this theory (which also applies to
secular resonances) is that the timescale to reach very high
eccentricity orbits may be long relative to the time taken to destroy
the comet by evaporation, limiting the ability of the resonance to
produce the wide range of pericentre distances that is inferred from the
models described above. Various solutions to this issue exist. One is
that some comets are larger than about 30~km in radius, so can survive
long enough to reach small pericentre distances
\citep{2000Icar..143..170B}. Another invokes the existence of additional
planets that either perturb the first planet, or perturb the comets
directly, thus causing larger orbit to orbit changes in the comet's
pericentre distances
\citep{1996Icar..120..358B,2000Icar..143..170B}. One of the ways that
this theory has been tested and developed is by aiming to match the
derived properties of FEBs, for example their distributions of
periastron distances, pericentre angles, and stellocentric distances at
transit \citep{2000Icar..143..170B}.

\section{Motivation}\label{s:orb}

Thousands of spectra of $\beta$ Pictoris have been taken in pursuit of
FEBs, but one aspect that has received little attention is the prospect
for directly deriving the cometary orbits. While orbital characteristics
have been derived based on the evaporation theory
\citep{1990A&A...236..202B,2014Natur.514..462K}, these are
model-dependent. That is, while the evaporation theory clearly
reproduces the observations very well, it necessarily includes many
assumptions for quantities that may vary from comet to comet, and it
would be far preferable to derive the orbits in a less model-dependent
way. Such an ability would provide further tests of the exocomet
hypothesis, and more certain tests of comet origin models. Orbital
constraints mean that the evaporation model could be refined, leading to
a better physical understanding of the comets and their comae.

To derive an orbit from the absorption lines, enough information to
constrain the orbital elements is needed (here we use pericenter
distance $q$, eccentricity $e$, inclination $i$, line of nodes $\Omega$,
longitude of pericentre $\varpi$, and true anomaly $f$). All six are not
necessary however, because we may assume an edge-on orbit, thus
eliminating $i$ and $\Omega$. We may further eliminate $f$ by defining
$\varpi$ to be the angle from the line of sight to the star at transit
centre (i.e.  $f = -\varpi$). An assumption, that the orbits are
parabolic ($e=1$) may be made based on the expectation that comets
cannot survive more than a few periastron passages near $\beta$ Pictoris
before evaporating. That is, the comets are not spiraling in on
near-circular orbits, but are appearing at just a few stellar radii via
relatively large orbit-to-orbit changes in $q$
\citep{1996Icar..120..358B}. At this point, two further independent
pieces of information are needed to constrain the orbit. In previous
work this has been the observed radial velocity of the absorption line,
and the distance to the star at transit estimated from the evaporation
model.

Given that FEBs are estimated to lie within a few tens of stellar radii
when they are observed, one may ask whether acceleration could be
detected in the course of a night's observations. For an object at
$10 R_\star$ from $\beta$ Pictoris (using $R_\star = 1.5R_\odot$ and
$M_\star = 1.7M_\odot$), the acceleration due to gravity is 2 m
s$^{-2}$, or 7~km~s$^{-1}$ h$^{-1}$. The magnitude of this acceleration
is within easy reach of high-resolution spectrographs such as the High
Accuracy Radial velocity Planet Searcher (HARPS), which has a spectral
resolution equivalent to a few km~s$^{-1}$.  Therefore, given that
absorption features inferred to be caused by absorption of material at
less than ten stellar radii are regularly seen, the expected
accelerations should be present and readily detectable in sequences of
spectra spanning more than about one hour.  Non-detection of
acceleration where sufficient temporal coverage exists would be a
serious issue for the exocomet hypothesis.

Aside from the expectation of acceleration given the likely orbits,
hints that it should be detectable can be seen in previous works;
\citet{1987A&A...185..267F} show a series of spectra taken over several
hours, in which a redshifted line clearly accelerates by about 5
km~s$^{-1}$, and temporal variation was also seen by
\citet{1996A&A...310..547L} and \citet{1999MNRAS.304..733P} With the
detection of such an acceleration, the distance derived from the
evaporation model would no longer be needed to constrain a parabolic
orbit (or could be used to refine the evaporation model).

\section{Detection of Exocomet Acceleration}\label{s:obs}

As a nearby (19pc) naked eye A6V star ($V=3.8$), $\beta$ Pictoris is
observable at high signal to noise ratios with a variety of
instruments. One particular focus has been high resolution
($\Delta \lambda/\lambda \sim 10^5$) spectra, which in addition to FEBs,
reveals a relatively stable circumstellar calcium absorption line
\citep[e.g.][]{1985ApJ...293L..29H} and stellar pulsations
\citep{2003MNRAS.344.1250K}, and yields constraints on the possible
orbits of unseen planets \citep{2006A&A...447..355G} and on the mass of
$\beta$ Pictoris b \citep{2012A&A...542A..18L}.

Specifically, the latter study obtained about 1000 HARPS spectra between
2004 and 2011. In the time since, observations have continued, and in
late 2017 all 2642 public HARPS spectra of $\beta$ Pictoris were
downloaded from the ESO archive for use in this study. These spectra
have been processed by the HARPS Data Reduction Software, and have been
shown by \citet{2014Natur.514..462K} to be of sufficient quality for
work related to FEBs; the spectra have a stable wavelength calibration
that far exceeds our needs here, and the relative flux calibration
across the spectral lines of interest and over the years since 2004 is
high enough that variations at the $\sim$1\% level are easily detected
(see for example their Extended Data Fig. 2).

\begin{figure}
  \begin{center}
    \hspace{-0.5cm} \includegraphics[width=0.5\textwidth]{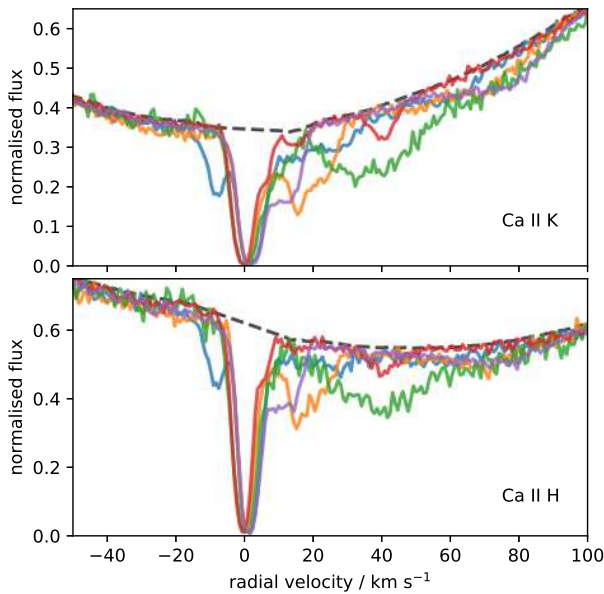}
    \caption{Examples of five HARPS spectra, showing the calcium K and H
      lines centred at the radial velocity of $\beta$ Pictoris. The
      dashed line shows the stellar reference spectrum. The absorption
      at zero km~s$^{-1}$ is thought to originate in a stable
      circumstellar disc, while the varying components are attributed to
      exocomet comae. Features at both low and high velocity are seen,
      with those at low velocity tending to be deeper.}\label{fig:ca}
  \end{center}
\end{figure}

Here, we focus on the calcium K \& H lines at 3933.66 and 3968.47
\AA. The only post-processing applied to the spectra here is to extract
the region within $\pm$500 km~s$^{-1}$ of the calcium lines, and apply a
multiplicative normalisation in the wings of the line so that all
spectra are on the same (arbitrary) flux scale. At this stage a
``stellar'' reference spectrum near each line is computed; the method
follows very closely that described by \citet{2014Natur.514..462K},
which used the highest flux values minus an estimated noise level across
all spectra to derive a spectrum free of absorption. An additional step
added here is interpolation across the circumstellar line, which is
always present but varies in width/velocity slightly, so that reference
divided (or subtracted) spectra do not show spurious variations that
might be interpreted as FEB activity near zero velocity.

An example of five ``randomly'' chosen spectra is shown in Figure
\ref{fig:ca}, where the stellar systemic velocity has been set to 21
km~s$^{-1}$.\footnote{i.e. the spectra shown were not chosen completely
  randomly, but only a few attempts need to be made to obtain a
  similarly varied set of spectra where both deep and blueshifted lines
  are present.} The obvious feature of all spectra is that the calcium
lines are broadened by a few hundred km~s$^{-1}$ by the fast stellar
rotation, and in addition every spectrum shows the stable circumstellar
line and several other narrow (few to a few tens of km~s$^{-1}$) and/or
broad (few tens to hundreds of km~s$^{-1}$) absorption features; the
FEBs. The stable circumstellar line is present in all spectra, and is
seen to move and/or change in width at the $\sim$km~s$^{-1}$
level. Spectra generally appear different night-to-night, because the
transit time for comets is of order hours. Absorption features are
generally present in both Ca lines, though are deeper and more apparent
in K than H unless the absorption is optically thick \citep[i.e. the
optical depth can be estimated from the line ratios,
e.g.][]{1992A&A...264..637L,1999MNRAS.304..733P}.

\begin{figure*}
  \begin{center}
    \hspace{-0.5cm} \includegraphics[width=1\textwidth]{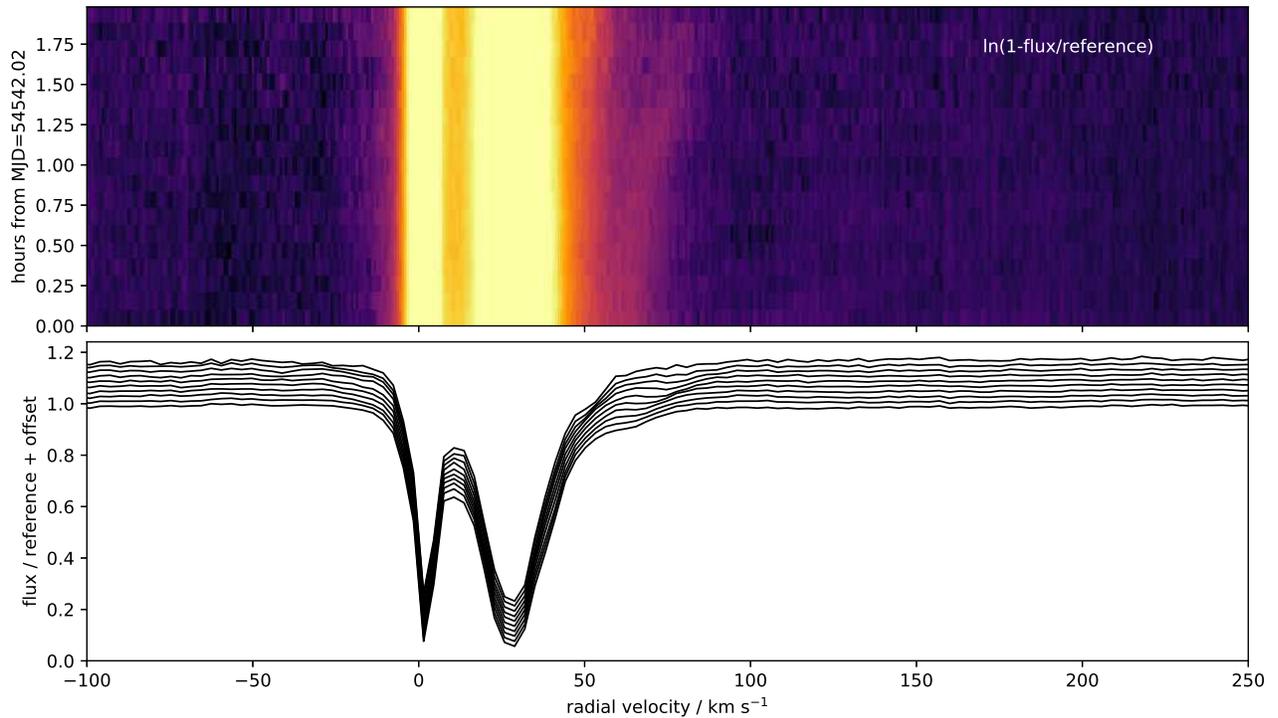}
    \caption{Example of an accelerating Ca K line absorption feature,
      seen at 60-80~km~s$^{-1}$. The \emph{upper panel} shows the level
      of absorption on a log stretch, and the \emph{lower panel} shows
      1d spectra that have been divided by the reference and offset
      vertically for clarity. The 1d spectra have been binned by a
      factor of two in time, and four in velocity, relative to the image
      in the upper panel. Two strong lines at $v_r \approx 0$ and 25
      km~s$^{-1}$, plus a weaker one moving from 60 to 80 km~s$^{-1}$,
      are visible, and the latter is the accelerating absorption
      feature. In the image this moving feature is seen as a faint
      stripe moving up and to the right over time. In the series of
      spectra the feature appears as a series of small absorptions that
      move up and to the right over time. Both panels illustrate clearly
      that this absorption feature is moving over the observation
      sequence.}\label{fig:eg}
  \end{center}
\end{figure*}

To detect accelerations we start with the graphical technique of
plotting temporal sequences of spectra as images, an example of which is
shown in the upper panel of Figure \ref{fig:eg}. The spectral dimension
is simply that output by the HARPS reduction pipeline, but spectra are
binned temporally; each vertical pixel is 1/250th of a day (about 6
minutes), so rows where multiple spectra fall are averaged. On nights
with continuous monitoring, bins typically contain five spectra. A weak
temporal variation is present on most nights, with a period of
approximately 0.6 hours; this has been removed by subtracting a smoothed
image with the absorption lines removed (see section \ref{s:fit}). A
final feature present is pulsation, whose primary signal is diagonal
stripes with a gradient of approximately 120 km~s$^{-1}$~h$^{-1}$; it is
strongest outside the deepest part of the lines ($\gtrsim$200
km~s$^{-1}$ at about the 1\% level) so not visible in Figure
\ref{fig:eg}, and not corrected for. The pulsation signal is much weaker
than the accelerating features considered here, and has a greater
velocity gradient than all fitted features, so is unlikely to affect the
results.

An alternative way to visualise the results is simply to plot the
spectra, as is done in the lower panel of Figure \ref{fig:eg}. This
panel shows the same spectra as in the upper panel, but binned
temporally by a factor of two, and spectrally by a factor of four.
These spectra can highlight temporal variations of absorption lines by
creating the illusion of a surface; static absorption lines result in a
series of features that are directly above each other (i.e. a vertical
`valley'), while those that accelerate show features that move in
velocity with each successive spectrum (i.e. a diagonal valley).

Figure \ref{fig:eg} shows three absorption features, two at 0 and 25
km~s$^{-1}$ that do not accelerate, and one at 60-80 km~s$^{-1}$ that
does. The accelerating feature is seen as either a stripe in the upper
panel, or a rightward-moving dip in the lower panel.

\begin{figure}
  \begin{center}
    \hspace{-0.5cm} \includegraphics[width=0.5\textwidth]{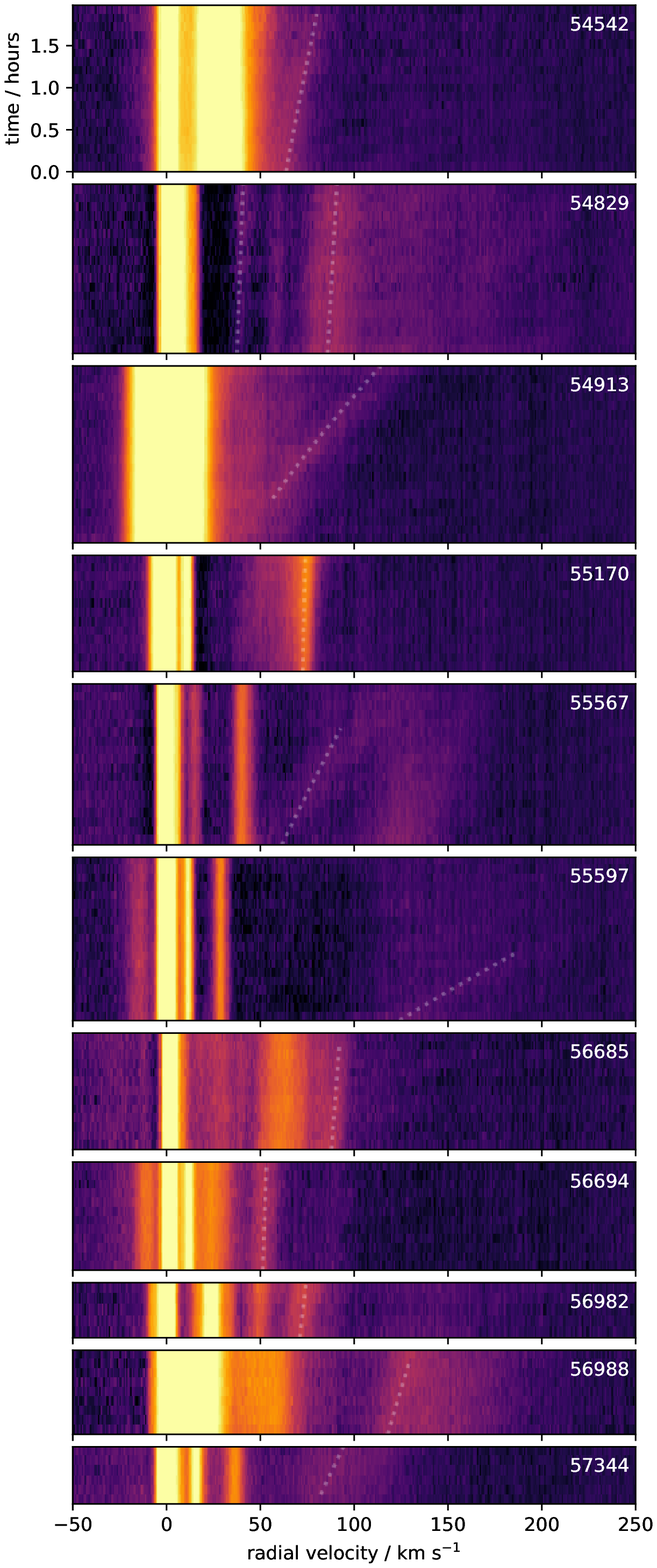}
    \caption{Accelerating features identified for modelling, indicated
      by transparent dashed lines. The modified Julian date at the start
      of each observing sequence is given. Each panel is the same as the
      upper panel in Figure \ref{fig:eg} and has the same scale, where
      lighter colours indicate stronger absorption.}\label{fig:all}
  \end{center}
\end{figure}

\begin{figure}
  \begin{center}
    \hspace{-0.5cm} \includegraphics[width=0.5\textwidth]{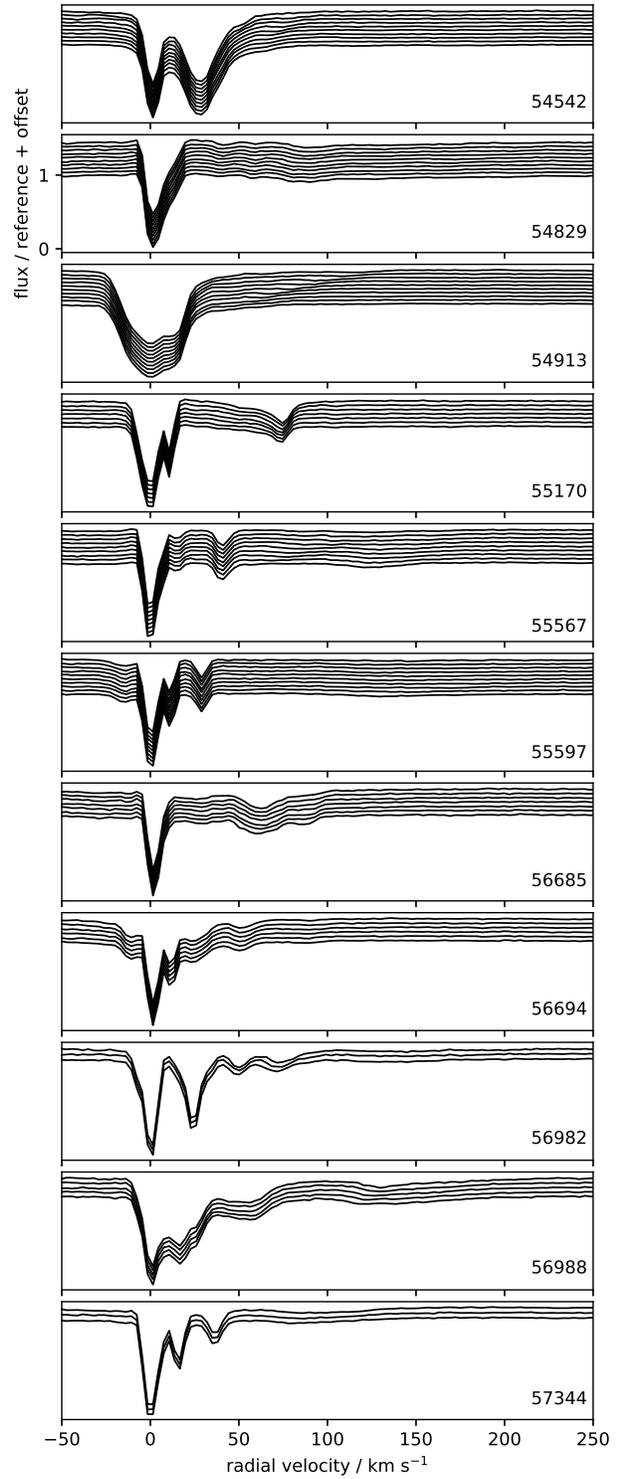}
    \caption{Spectra of accelerating features identified for
      modelling. The modified Julian date at the start of each observing
      sequence is given. Each panel is the same as the lower panel in
      Figure \ref{fig:eg}. As discussed in the text, weak accelerating
      features are not always easily discernible.}\label{fig:all_spec}
  \end{center}
\end{figure}

A set of accelerating features were identified in spectra by eye using
images similar to Figure \ref{fig:eg}, from 35 nights' data with more
than about an hour of continuous observation, and these are shown in
Figures \ref{fig:all} and \ref{fig:all_spec}. The dates on which these
sequences started are shown in each panel, and given in Table
\ref{tab:elem}. In most cases the accelerating feature is apparent,
though the relatively low contrast relative to the deeper circumstellar
line makes their visualisation difficult in some cases.  Also, in a few
cases the acceleration is not large enough to be clearly visible
(e.g. the 40 km~s$^{-1}$ feature in second panel from top is
accelerating slightly, but significantly). While not all features are
clearly visible in the figures, the identified features are verified to
i) have average absorption significantly greater than zero (i.e. exist),
and ii) be accelerating significantly, by the fitting method outlined in
section \ref{s:fit}.

All accelerating features found are red shifted, and are accelerating
towards the star. No automated detection methods were attempted, though
we found that some features could also be discovered by fitting
Gaussians to absorption features in individual spectra, and plotting the
temporal evolution of their radial velocities \citep[see][for such
plots]{1999MNRAS.304..733P}. In addition, distinct unblended absorption
features that did not show apparent acceleration were also selected for
further analysis, with the expectation that useful orbital constraints
could still arise (e.g. a minimum distance at transit). For these, only
red shifted events were selected because the aim is to illustrate
orbital fitting of accelerating features, and to compare the orbital
constraints for both accelerating and non-accelerating features. There
is a degree of subjectivity in the subset of spectra used below, and
more accelerating features might be discovered with a more systematic
approach. We are not however attempting to make any statistical
inference from the subset of features analysed, so any biases introduced
by our approach do not influence the results that follow.

\section{Orbit Fitting}\label{s:fit}

Given an accelerating absorption feature it is relatively simple to
construct and fit a model of an orbit. The primary assumption here is
that the orbit crosses the centre of the stellar disc, leaving three
elements to be constrained (eccentricity $e$, longitude of pericenter
$\varpi$, and pericentre distance $q$). Thus, an orbital fit will yield
constraints, but with a strong degeneracy between these within the
allowed parameter space. However, as discussed above the reasonable
assumption of a parabolic orbit ($e=1$), means that an accelerating
feature contains enough information to constrain the orbit fully. Here
we illustrate this degeneracy by leaving $e$ to vary, but also restrict
the results to parabolic orbits.

Radial velocity $v_r$ is the variable being fitted here, with respect to
Earth once $\beta$ Pictoris' systemic velocity has been subtracted (i.e.
$v_r$ is only the same as the stellocentric radial velocity at transit
center). This velocity can be derived using the motion of an orbit in a
frame $\dot{x}, \dot{y}$, where the pericentre lies along the positive
$x$ coordinate, and $\beta$ Pictoris is at the origin. The system is
viewed from the negative $Y$ direction, where the $X$, $Y$ coordinates
are rotated clockwise by $\varpi$ from $x$, $y$
\cite[e.g.][]{2000ssd..book.....M}. That is, the orbit in the $x,y$
frame is rotated such that the pericentre direction is at an angle
$\varpi$ from $-Y$ (the reader may wish to refer ahead to Fig.
\ref{fig:orb}.)  In the frame of the orbit
\begin{align}
  \dot{x} &= - \mu/h \sin f \nonumber \\
  \dot{y} &= \mu/h (1 + \cos f)
\end{align}
where $\mu = GM_\star$, $f$ is the true anomaly, $h^2= \mu p$, and
$p=2q$ for parabolic orbits and $p=a(1-e^2)$ for elliptical orbits. The
rotation yields
\begin{align}\label{eq:rv}
  v_r = \dot{Y} &= -\dot{x} \cos{\varpi} + \dot{y} \sin{\varpi} \nonumber \\
                &= \mu / h \left( \sin f \cos \varpi + \sin \varpi [ 1+ \cos f ] \right)
\end{align}

We fit absorption features directly (i.e. instead of fitting
accelerations and then deriving orbital constraints), so deriving an
orbit from the images requires five or six free parameters, the orbital
elements $q$, $\varpi$, and $e$ (for elliptical orbits), the time $t_0$
of mid-transit relative to the start of the observation sequence, and
the (constant) depth $\delta$ and velocity width $\sigma$ of the
absorption line, which is assumed to be Gaussian. The radial velocity at
a given time $t$ is found by first finding the true anomaly $f(t)$
(using Barker's equation in the case of parabolic orbits), and then
using equation (\ref{eq:rv}). Thus, a model of an absorption line
analogous to the accelerating feature in Figure \ref{fig:eg} is created
by computing $v_r(t)$ at the centre of each temporal bin, and the image
comprises a vertical sequence of (negative) Gaussians of depth $\delta$
(where positive values mean absorption) and width $\sigma$ centred at
$v_r(t)$. At each time, an absorption line is only assumed to be present
if the centre of the body is in front of the star (i.e.
$|X| < R_\star$). Additional parameters that can be derived from the
models\footnote{We do not distinguish between $d_{\rm tr}$ and
  $d_{\rm tr, mid-transit}$ because the distance to the star changes
  little during a transit. We use $\dot{v}_r$ instead of
  $\dot{v}_{r, {\rm mid-transit}}$ for the same reason.} are the radial
velocity at mid-transit $v_{r, {\rm mid-transit}}$, the distance to the
star at mid-transit $d_{\rm tr}$ ($=-Y$), and the acceleration
$\dot{v}_r = \mu / d_{\rm tr}^2$.

As can be seen from Figures \ref{fig:all} and \ref{fig:all_spec}, most
spectra contain fixed absorption lines that are much deeper than the
accelerating ones. Fitting the model and obtaining reasonable $\chi^2$
values from residual images that are also easily interpreted visually
therefore requires that these are removed.  First, we find the median
spectrum over the nightly sequence. The noise level in each temporal bin
is then computed as the standard deviation of the residuals when this
median is subtracted from the initial image. Next, a second order
polynomial is fitted to the median in the region near the line of
interest (typically 20 km~s$^{-1}$ on either side, though in cases with
other nearby lines (e.g. for lines with low $v_r$) this region is
necessarily smaller). The outer edges of this region have twice the
weight relative to the centre in the fit, and this polynomial is taken
as an estimate of the local time-independent background near the
accelerating absorption feature. This polynomial fit to the median was
found to be much better than simply using the median, as the absorption
line of interest was commonly partially subtracted with the latter
method. During the fitting we subtract this polynomial and the proposed
model from the image to obtain a residual image.

Fitting orbits to the spectra is done in two steps, where the goodness
of fit metric is the sum of squares of the residuals divided by the
noise. First, a model is fitted to the original image with the
polynomial subtracted to obtain a preliminary fit. A preliminary
residual image is created by subtracting the polynomial and this model,
and the residuals are smoothed with a small temporal extent of 8.5
minutes, but a wide spectral extent of 100 km~s$^{-1}$. These smoothed
residuals (which vary at the percent level relative to the absorption
features) are then subtracted from the original image before fitting.

As noted above, very strong degeneracies between fitted parameters exist
when the eccentricity is not fixed, so the second step is to use a
Markov Chain Monte Carlo (MCMC) method to explore the parameter space.
Specifically, we use the ensemble sampler \texttt{emcee}
\citep{gw2010,2013PASP..125..306F} with 256 parallel chains
(``walkers''). The chains are run for 100,000 steps to ensure the
parameter space is fully explored, as in some cases the autocorrelation
lengths were of order 10,000 steps. The only restrictions on the fitted
parameters are that orbits are bound or parabolic (i.e.  $e \le 1$), do
not hit the star at pericentre (i.e.  $q > R_\star$), and that the depth
$\delta$ and width $\sigma$ are positive.

After these runs, the walkers from the final step were used to generate
distributions of orbital elements and other derived quantities of
interest.  This fitting was run twice; once for elliptical orbits, and
then again for parabolic orbits, thus obtaining two sets of orbital
elements from each modelled absorption line. The parabolic orbits are of
course a subset of the elliptical orbits, but fitting them is more
satisfactory than simply applying an arbitrary cut in eccentricity to
the elliptical orbit results (at $e=0.95$, say).

\begin{figure}
  \begin{center}
    \hspace{-0.5cm} \includegraphics[width=0.5\textwidth]{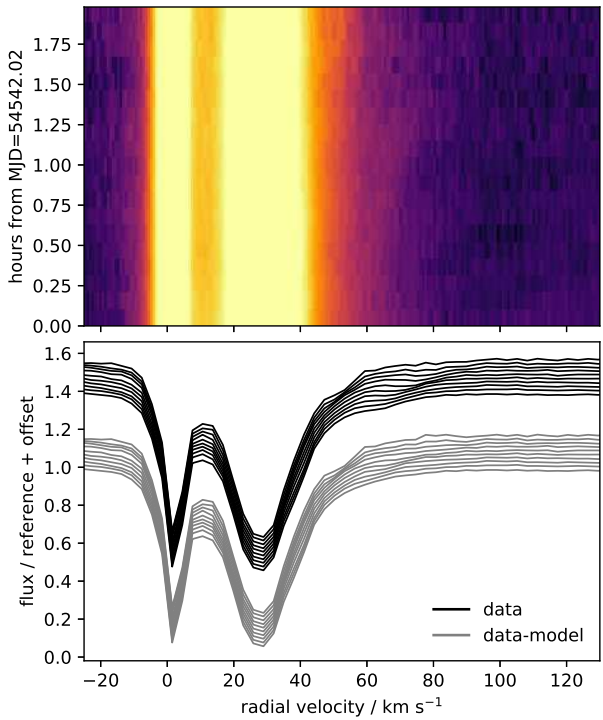}
    \caption{Fitting results, showing the same data as in Figure
      \ref{fig:eg}, but with the best-fit model subtracted. The
      \emph{upper panel} shows the level of absorption after model
      subtraction, where brighter colours indicate deeper
      absorption. The \emph{lower panel} shows the binned spectra, with
      the black lines showing the original data, and the grey lines the
      model subtracted data.}\label{fig:fit}
  \end{center}
\end{figure}

An example of a model fit is shown in Figure \ref{fig:fit}, for the same
feature in Figure \ref{fig:eg} (and the top row of Figure
\ref{fig:all}). While the accelerating feature is not subtracted
perfectly, the model adequately reproduces the accelerating absorption
line, and while the depth of the absorption can appear to vary
\citep{1994A&A...282..804B}, any variation is not at a sufficiently
strong level that the addition of another model parameter is warranted
for the purposes of orbital fitting. This procedure was applied to all
eleven sets of spectra shown in Figure \ref{fig:all}, plus others where
non-accelerating lines were well-separated from others, thus obtaining
sets of elliptical and parabolic orbital elements from a set of 27
absorption features. Of these, twelve are accelerating features (there
are two in the second panel from the top in Figure \ref{fig:all}).

\begin{figure}
  \begin{center}
    \hspace{-0.5cm} \includegraphics[width=0.5\textwidth]{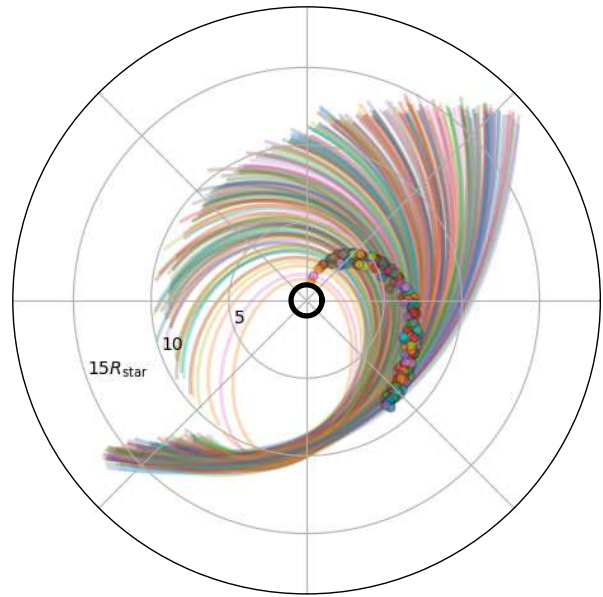}
    \caption{Distribution of possible elliptical orbits for the example
      in Figure \ref{fig:eg}. Each line shows one of the 256 walkers at
      the final step in the MCMC chains, and each has a corresponding
      circle at pericenter. The axes here are $X$, $Y$, so pericentre is
      at an angle $\varpi$ anticlockwise from $-Y$.  The thick black
      circle in the centre is the star, and the system is viewed from
      below (i.e. $-Y$).}\label{fig:orb}
  \end{center}
\end{figure}

\section{Results}\label{s:res}

The results of the orbital fits may be visualised in various ways;
Figure \ref{fig:orb} shows a top-down ($X$, $Y$) view of the final set
of orbits for the example above. From this figure the degeneracies are
clear. A wide range of orbits is possible, and those with smaller
pericentre distances reach pericentre at a greater longitude from the
line of sight to the star. The degeneracy with eccentricity is more
subtle and harder to discern here, as the most eccentric orbits are in
fact those with the greatest pericentre distances and smallest
pericentre longitudes.

\begin{figure}
  \begin{center}
    \hspace{-0.5cm} \includegraphics[width=0.5\textwidth]{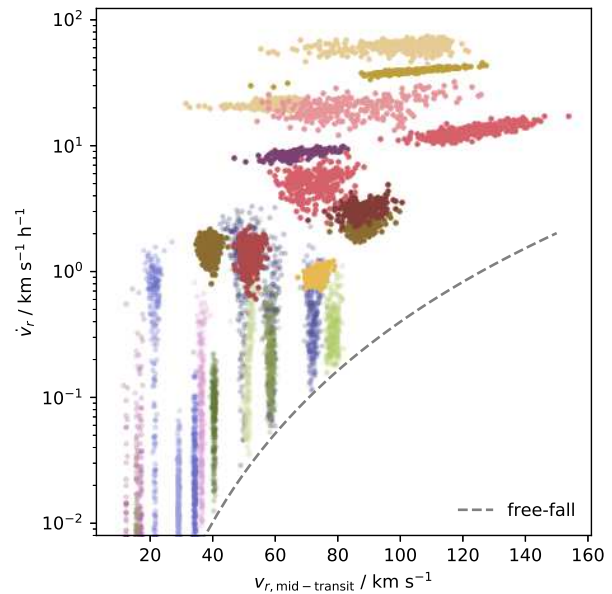}
    \caption{Accelerations for fitted absorption features, plotted
      against the distance to the star at mid-transit. Each cluster of
      dots shows the final step of the 256 walkers in the MCMC chain for
      the fit to a given absorption feature. The dot colours are used
      consistently across Figs. \ref{fig:d_acc}-\ref{fig:dep_d}. The
      dashed line shows the acceleration expected for bodies in
      free-fall from infinity. Those without detected acceleration
      extend down to the free-fall line, and their symbols are
      transparent.}\label{fig:d_acc}
  \end{center}
\end{figure}

These orbits may also be viewed in parameter space plots, each of which
shows clusters of 256 points from the final step in the MCMC
chains. These clusters therefore show the distribution of possible
parameters for a given absorption feature. The first example is Figure
\ref{fig:d_acc}, which illustrates the distinction between absorption
features that do and do not show measurable accelerations. While
features with significant acceleration yield a direct measurement of the
distance to the star, features with undetectable accelerations are
consistent with orbits that pass in front of the star at arbitrarily
large distances. However, the restriction that fitted orbits are bound
or parabolic means that for a given velocity at mid-transit the smallest
possible acceleration coresponds to a object in free-fall from infinity
(which is $\dot{v}_r=v_r^4/(4\mu)$). Thus, non- accelerating features
are easily identified in Figure \ref{fig:d_acc} by their lack of
clustering at a given $\dot{v}_r$, and the extension of their
distributions down to the expected limit based on the fitting method.

\begin{figure*}
  \begin{center}
    \hspace{-0.5cm} \includegraphics[width=0.5\textwidth]{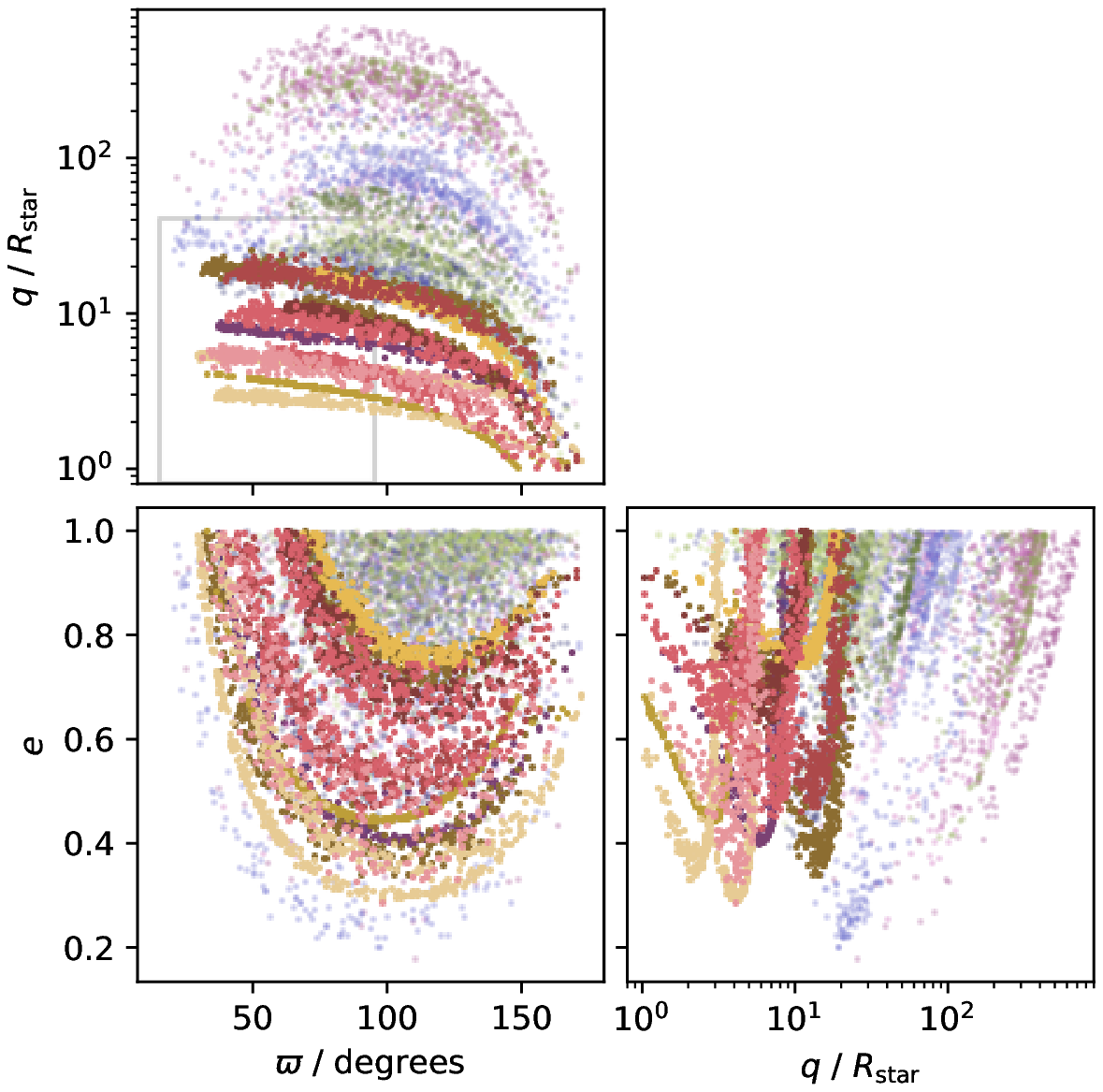}
    \hspace{-0.cm} \includegraphics[width=0.5\textwidth]{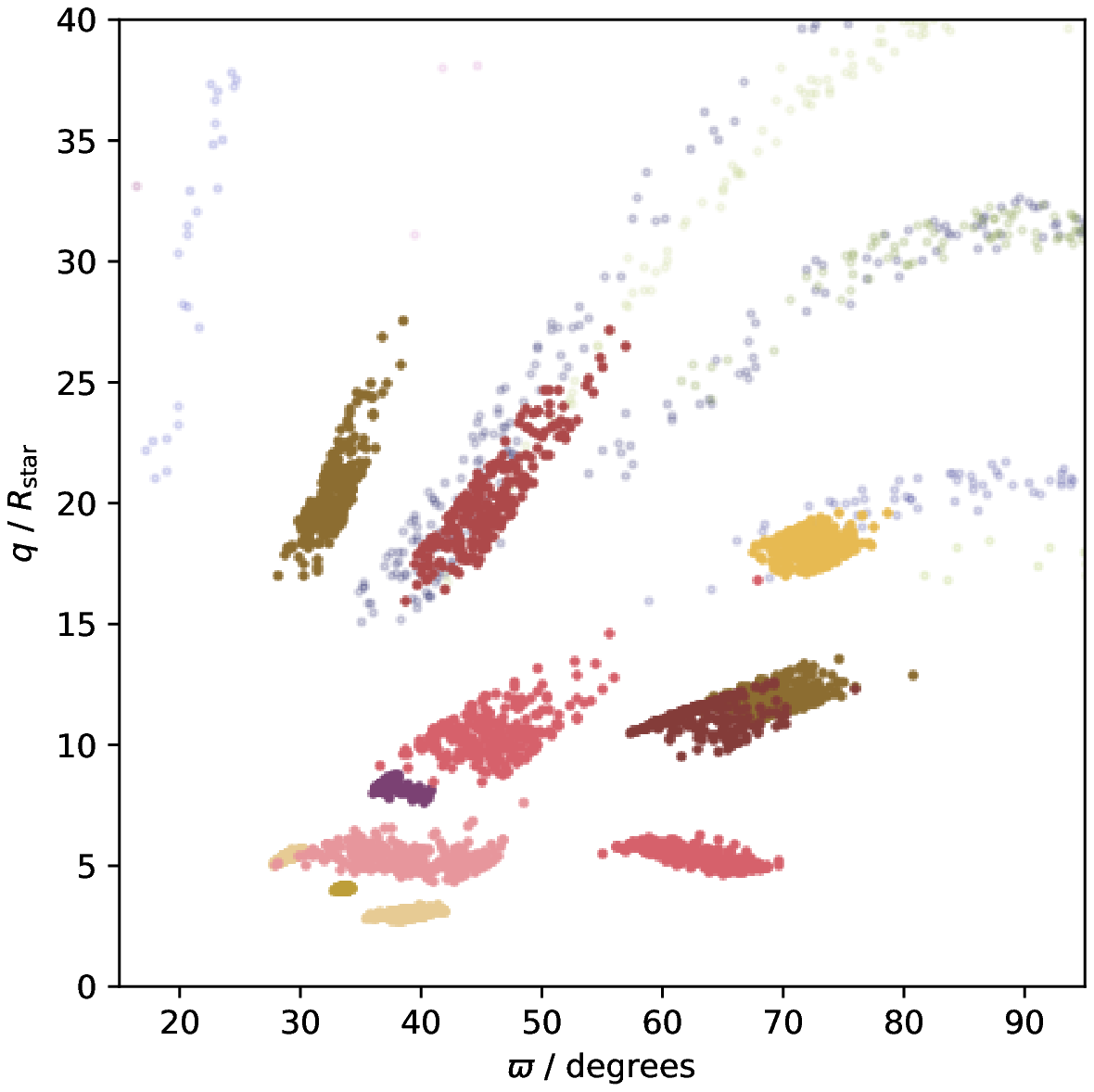}
    \caption{Distributions of possible orbital elements for all fitted
      features. Each cluster of dots shows the final step of the 256
      walkers in the MCMC chain for the fit to a given absorption
      feature. Those without detected acceleration are transparent and
      dot colours are used consistently across
      Figs. \ref{fig:d_acc}-\ref{fig:dep_d}. The \emph{left panel} shows
      results for elliptical orbits, and the \emph{right panel} shows
      results for parabolic orbits. The range in the right panel is
      restricted (to the grey box in left panel), as the poorly
      constrained parabolic orbits look very similar to the elliptical
      ones. For elliptical orbits the orbital element degeneracies are
      strong, but the elements are well constrained for parabolic orbits
      when acceleration is seen.}\label{fig:elem}
  \end{center}
\end{figure*}

In the left panel of Figure \ref{fig:elem}, all fitted orbits are shown,
and again the distinction is made between orbits with and without
detected acceleration. It is clear that degeneracies apply in all
cases. The same trends visible in Figure \ref{fig:orb} can be
identified, but it is now apparent that each fit reaches a minimum
eccentricity that is consistent with the data. This can be understood
from equation (\ref{eq:rv}), because at transit $f \approx -\varpi$, and
therefore $v_{r, {\rm transit}} \propto \sin \varpi$ and the maximum
radial velocity at transit for a given orbit occurs when $\varpi$ is
near 90$^\circ$. Thus, the lowest eccentricity orbit consistent with the
data should also be near 90$^\circ$. The eccentricities rise to larger
pericentre longitudes, but stop short of $e=1$ because of the
restriction that the pericentre distance be larger than $R_\star$. In
cases with detected acceleration $e$ and $\varpi$ are strongly
correlated, but this correlation is not present for those without and
the distributions form a broad cloud of points centered near
$\varpi \approx 100^\circ$.

Restricting the orbits to be parabolic yields tighter constraints, as
shown in the right panel of Figure \ref{fig:elem}. The plotted range is
smaller here, because non-accelerating features yield very similar
constraints to those for elliptical orbits. For accelerating features
however, the constraints localise the orbits to relatively small regions
in $q - \varpi$ space. A summary of the parameters derived for parabolic
orbits is given in Table \ref{tab:elem}.

\begin{figure*}
  \begin{center}
    \hspace{-0.5cm} \includegraphics[width=0.5\textwidth]{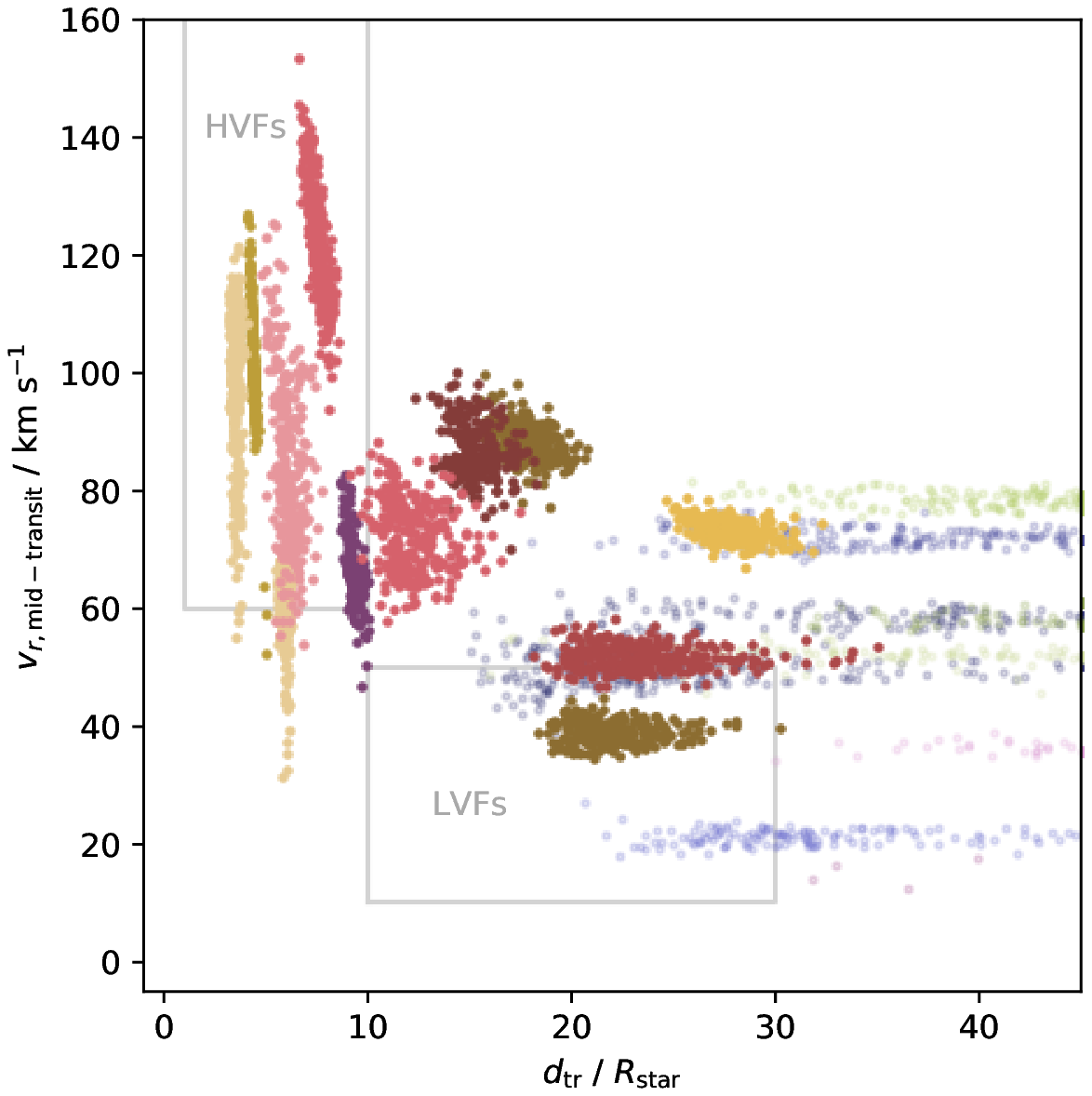}
    \hspace{-0.5cm} \includegraphics[width=0.5\textwidth]{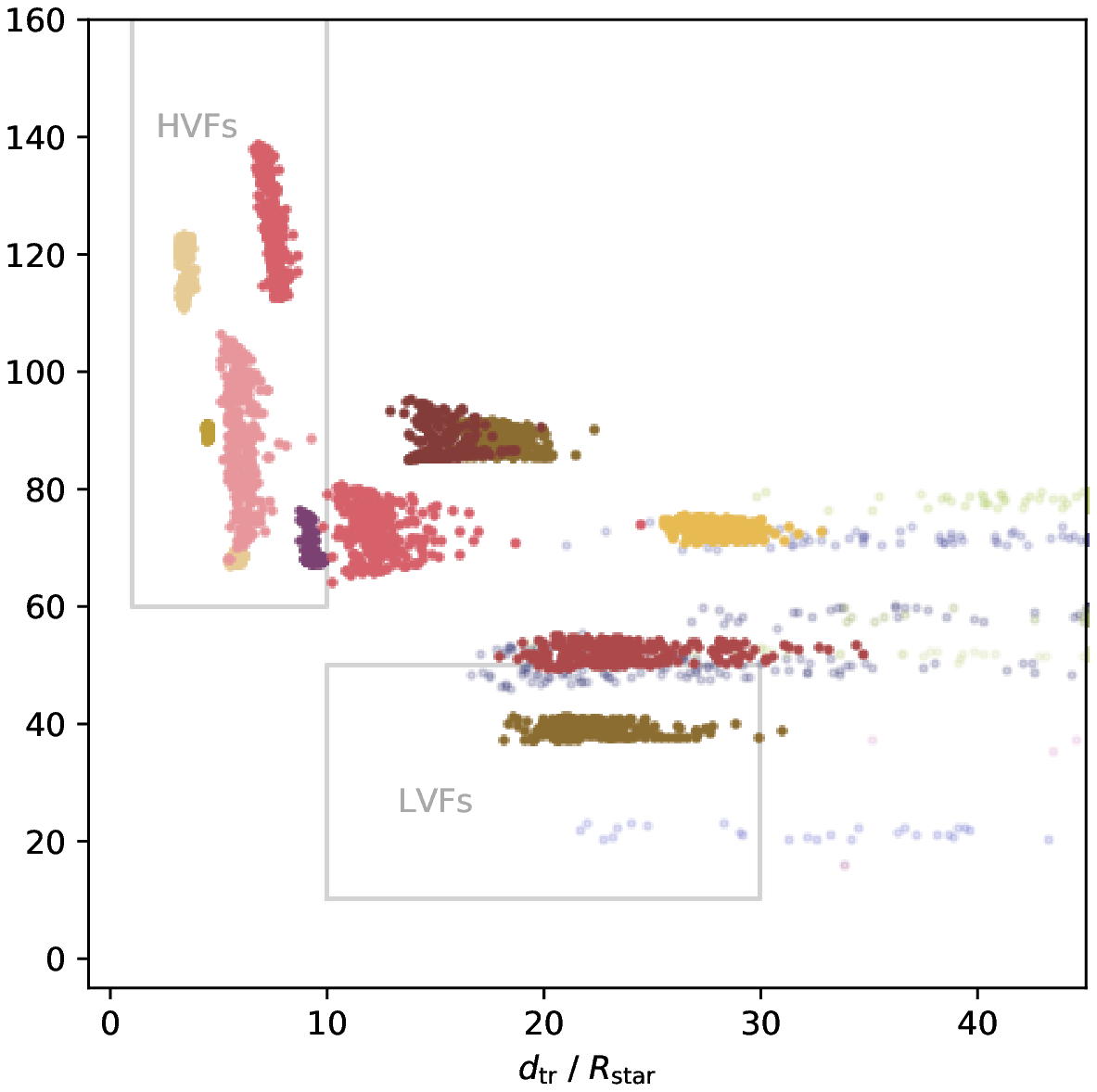}
    \caption{Distributions of velocity versus stellocentric distance at
      mid-transit, for elliptical (\emph{left panel}) and parabolic
      (\emph{right panel}) orbits. Each cluster of dots shows the final
      step of the 256 walkers in the MCMC chain for the fit to a given
      absorption feature. Those without detected acceleration are
      transparent and dot colours are used consistently across
      Figs. \ref{fig:d_acc}-\ref{fig:dep_d}. The grey boxes show `high'
      and `low velocity features', following
      \citet{2000Icar..143..170B}. The constraints in this space are
      only moderately improved when parabolic orbits are
      assumed.}\label{fig:rv_d}
  \end{center}
\end{figure*}

A final way to view the orbit distributions is shown in Figure
\ref{fig:rv_d}, primarily for comparison with the models presented in
\citet{2000Icar..143..170B} (e.g. see their Figure 4). Here the radial
velocity at the time of mid-transit is used; while Figure \ref{fig:all}
shows that in general the velocity of an accelerating feature at any
given time is well constrained, this is not necessarily true at the time
of mid-transit (which may occur during, before, or after the observation
sequence). Thus, as shown in the left panel, for elliptical orbits the
distribution of $v_{r, {\rm mid-transit}}$ can be rather wide. When the
orbits are restricted to be parabolic, the right panel shows that the
constraints become tighter, though not so much that this assumption
yields significantly more information than when eccentricity is a free
parameter. That is, while parabolic orbits must be assumed to obtain
tight constraints on orbital elements, this is less true in the
parameter space of Figure \ref{fig:rv_d}.

Non-accelerating features result in poor constraints in Figure
\ref{fig:rv_d}, most of which extend beyond the right side of the
plot. The distances at transit are only constrained to be larger than
some value that depends on the strength of the signal (i.e. the upper
limit on the level of acceleration that was not detected). However, the
fact that the absorption features are seen at all means that
$d_{\rm tr}$ is no more than a few tens of stellar radii for evaporation
to be occuring \citep[][derive $\sim$35$R_\star$]{1996A&A...310..181B},
so the true $d_{\rm tr}$ of these orbits is likely below this value, and
consistent with the orbits with stronger constraints.

In Figure \ref{fig:rv_d} the general trend is that orbits that transit
at smaller stellocentric distances have a greater radial velocity. The
grey boxes show the estimated regions in which previously identified
populations of FEBs lie, where the evaporation model was used to
estimate the distance at transit, and the distinction between ``high''
and ``low velocity features'' (HVFs and LVFs) is largely
historic.\footnote{Here the `very low velocity features' (VLVFs) are not
  considered, as none in this class were found to be accelerating.} The
first conclusion is therefore that the orbits derived here in general
have similar properties to those derived using the evaporation model.

While our orbits tend to lie within the HVF box, they are on average
above the LVFs. However, the lack of orbits falling within the LVF box
is probably the result of a bias, as features that are the result of
transits at greater distances more commonly occur at low velocity, and
these are more common and more likely to be blended with other
features. Thus, the main conclusions from placing the orbits derived
here on this plot is that there is not strong evidence for a lack of
orbits within the HVF and LVF boxes, but that orbits certainly exist
outside them, in particular at greater velocities for a given distance
at transit when $d_{\rm tr} > 10 R_\star$. The existence of such orbits
is indeed predicted by the resonance model \citep{2000Icar..143..170B},
and the orbits derived from this method will provide useful new
constraints for this model.

\begin{figure}
  \begin{center}
    \hspace{-0.5cm} \includegraphics[width=0.5\textwidth]{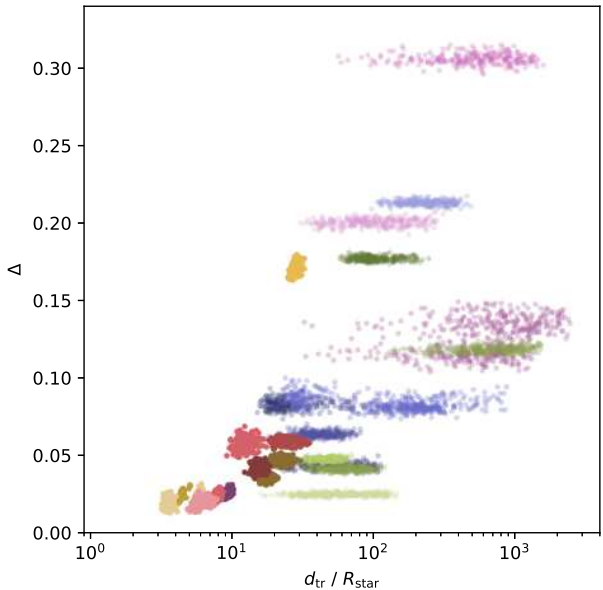}
    \caption{Correlation between distance at transit and absorption
      fraction for elliptical orbits. Each cluster of dots shows the
      final step of one of the 256 walkers in the MCMC chain for the fit
      to a given absorption feature. Those without detected acceleration
      are transparent and dot colours are used consistently across
      Figs. \ref{fig:d_acc}-\ref{fig:dep_d}. This link is predicted by
      the evaporation model because exocomet comae are more strongly
      ablated closer to the star, and therefore cover a smaller fraction
      of the star. The constraints are not improved with the assumption
      of parabolic orbits (so that plot is omitted
      here).}\label{fig:dep_d}
  \end{center}
\end{figure}

The last view of the fitting results includes the absorption depth
derived in each case, shown against the distance at transit in Figure
\ref{fig:dep_d}. The values for $\delta$ are in units of normalised flux
(i.e. as in Figure \ref{fig:ca}), but can be converted to fractional
peak absorption $\Delta$ using the reference spectrum with
$\Delta = \delta / F_{\rm ref}$, where $F_{\rm ref}$ is the reference
(stellar) flux at $r_{v, {\rm mid-transit}}$. The relevance of this plot
is that the evaporation model described above predicts that comet
absorption is deeper for more distant transits, because the effect of
stellar radiation pressure on the calcium ions is less. Thus the comet
comae are larger and cover a greater fraction of the projected stellar
surface. This prediction is borne out by Figure \ref{fig:dep_d}, which
shows a clear tendency for the closest transits to have shallower
absorption. As with Figure \ref{fig:rv_d}, the likely distances for
poorly constrained orbits are towards the low $d_{\rm tr}$ ends of their
distributions, so the true correlation may be as tight as suggested by
the clouds of points for orbits derived from accelerating features.

\section{Discussion and Conclusions}

This work is motivated by the realisation that previously estimated
orbital elements for $\beta$ Pictoris' comets imply acceleration of the
absorption features should be visible in spectra taken over several
hours. Figures \ref{fig:eg}, \ref{fig:all} and \ref{fig:all_spec} show
that these accelerations are indeed seen, and using a simple model these
accelerating features can be used to derive orbital elements for
individual comets. The constraints on the pericentre distance $q$ and
longitude $\varpi$, and eccentricity $e$ are degenerate, but constraints
on the distance to the star at the time of transit are less so. Thus,
useful constraints that can be directly compared to models of the
comets' origins can be derived in a model-independent way.

The constraints can be further improved with the reasonable assumption
that the orbits are parabolic, because comets on low eccentricity orbits
spend too much time near the star and may not survive long enough to be
observed. With this assumption, the pericentre distance and longitude
are well constrained, and the latter lies in the range 25 to 75$^\circ$
for the orbits modelled here (Fig. \ref{fig:elem}). That is, the
tendency for the pericentre longitudes to lie in a restricted range
suggested in the past based on orbits estimated using an evaporation
model \citep[e.g.][]{1990A&A...236..202B} is also found when the orbits
are derived directly.

A further test of the evaporation model is of the prediction that comet
absorption is deeper for more distant transits, because comae of closer
comets are more strongly ablated and therefore cover less of the stellar
surface. This correlation is also borne out by our analysis
(Fig. \ref{fig:dep_d}).

Our technique of exocomet orbit fitting is analogous to other techniques
that measure the radial velocity of planets directly, by using
cross-correlation with models of CO and H$_2$O absorption to show that
their spectra are double-lined
\citep[e.g.][]{2010Natur.465.1049S}. Because they can measure the planet
velocity with a precision of a few percent, rather than the reflex
motion of the host star, these techniques yield stringent orbital
constraints, for example on the eccentricity and inclination of
non-transiting planets \citep{2012Natur.486..502B}. These observations
have detected both absorption in the thermal emission spectrum of the
planet \citep[e.g.][]{2017AJ....153..138B}, and absorption in
transmission spectra (i.e. during transit)
\citep[e.g.][]{2010Natur.465.1049S,2018arXiv180109569B}. The latter
detections are analogous to the results presented here, and future
transmission spectroscopy of transiting planets may also yield
independent and useful constraints on the orbital elements.

Our results are reasonably straightforward because the approach is
simple. However, a few caveats and inbuilt assumptions should be noted.
In fitting orbits, the stellar radius and mass have been assumed fixed,
so there remains a small systematic uncertainty in the results. We
further assumed that the transits cross the centre of the stellar disc,
but there are no cases where both ingress and egress are seen so this
assumption should not influence the results significantly. In some cases
the absorption lines are probably blended, meaning that a single model
may have been applied where two might have been more appropriate; the
most likely examples are the 80 km~s$^{-1}$ feature in the second panel
and the two broad features between 50 and 150~km~s$^{-1}$ in the fifth
panels in Figures \ref{fig:all} and \ref{fig:all_spec}. Again, the
results are unlikely to be significantly changed by use of a more
complex model, but the uncertainties on the derived orbital parameters
may increase. The most important issue for further work to address is
probably a systematic and objective identification of accelerating
features. Other issues include i) simultaneous fitting of H and K lines
where possible, ii) simultaneous fitting of possibly bended lines, and
iii) modelling variable absorption during transit.

This work has various possible implications. Most obviously, the
accelerating absorption features provide strong evidence that the FEB
phenomenon indeed originates in bodies that pass in front of $\beta$
Pictoris at close distances (for which comets is the leading
hypothesis). By fitting models to these features the orbital properties
are found to be consistent with previous estimations that were based on
an evaporation model, and therefore these results largely validate that
model as a reasonable physical explanation of the comet comae. Further
work could attempt to refine the evaporation model by attempting to
again fit the absorption features, but now using orbital element
constraints derived when acceleration is detected.

\section*{Acknowledgments}

GMK is supported by the Royal Society as a Royal Society University
Research Fellow. I am grateful to Matteo Brogi and Daniel Bayliss for
discussions during the preparation of this work, and to the referee for
a valuable review.

The code used in this research is available on github at
\href{https://github.com/drgmk/feb-accel}{https://github.com/drgmk/feb-accel},
including the final steps of the chains used in Figures \ref{fig:orb} to
\ref{fig:dep_d}.


\appendix

\begin{table*}
  \begin{center}
    \caption{Median parameters derived for accelerating features
      assuming parabolic orbits, and their standard deviations. MJD
      refers to the modified Julian date on which the spectral
      observations began, and the table is listed in the same order as
      Figure \ref{fig:all}.}\label{tab:elem}
    \begin{tabular}{lrlrlrlrlrlll}
      \hline
      \multicolumn{1}{c}{MJD} & \multicolumn{1}{c}{$\dot{v}_r$} & \multicolumn{1}{c}{$\sigma_{\dot{v}_r}$} & \multicolumn{1}{c}{$v_r$} & \multicolumn{1}{c}{$\sigma_{v_r}$} & \multicolumn{1}{c}{$q$} & \multicolumn{1}{c}{$\sigma_q$} & \multicolumn{1}{c}{$\varpi$} & \multicolumn{1}{c}{$\sigma_{\varpi}$} & \multicolumn{1}{c}{$d_{tr}$} & \multicolumn{1}{c}{$\sigma_{d_{tr}}$} & \multicolumn{1}{c}{$\Delta$} & \multicolumn{1}{c}{$\sigma_\Delta$} \\
      \multicolumn{1}{c}{d} & \multicolumn{2}{c}{km s$^{-1}$ h$^{-1}$} &
                                                                         \multicolumn{2}{c}{km s$^{-1}$} & \multicolumn{2}{c}{$R_\star$} & \multicolumn{2}{c}{rad} & \multicolumn{2}{c}{$R_\star$} & \multicolumn{2}{c}{\%} \\
      \hline
54542 & 9 & 0.4 & 68 & 2.0 & 8 & 0.2 & 0.6 & 0.02 & 9 & 0.2 & 0.03 & 0.001 \\
54829 & 2 & 0.3 & 88 & 1.8 & 12 & 0.5 & 1.2 & 0.04 & 18 & 1 & 0.04 & 0.0009 \\
54829 & 2 & 0.3 & 39 & 1.1 & 20 & 2 & 0.6 & 0.03 & 22 & 2 & 0.05 & 0.001 \\
54913 & 38 & 0.7 & 90 & 0.6 & 4 & 0.04 & 0.6 & 0.004 & 4 & 0.04 & 0.03 & 0.0005 \\
55170 & 1 & 0.1 & 73 & 1.2 & 18 & 0.5 & 1.3 & 0.03 & 28 & 1 & 0.2 & 0.002 \\
55567 & 22 & 1.1 & 68 & 0.5 & 5 & 0.1 & 0.5 & 0.008 & 6 & 0.1 & 0.02 & 0.0008 \\
55597 & 65 & 6.0 & 117 & 3.3 & 3 & 0.1 & 0.7 & 0.02 & 3 & 0.2 & 0.02 & 0.001 \\
56685 & 3 & 0.4 & 86 & 2.8 & 11 & 0.5 & 1.1 & 0.05 & 15 & 1 & 0.04 & 0.002 \\
56694 & 1 & 0.3 & 52 & 1.4 & 20 & 2 & 0.8 & 0.06 & 23 & 3 & 0.06 & 0.002 \\
56982 & 5 & 1.0 & 73 & 4.1 & 10 & 1 & 0.8 & 0.06 & 12 & 1 & 0.06 & 0.003 \\
56988 & 13 & 1.3 & 125 & 7.3 & 5 & 0.4 & 1.1 & 0.06 & 7 & 0.4 & 0.02 & 0.002 \\
57344 & 21 & 3.4 & 89 & 9.7 & 5 & 0.5 & 0.7 & 0.07 & 6 & 0.5 & 0.02 & 0.002 \\
     \hline
    \end{tabular}
  \end{center}
\end{table*}

\end{document}